# Arbitrary synthetic dimensions via multi-boson dynamics on a one-dimensional lattice

Dali Cheng,[1] Bo Peng,[1] Da-Wei Wang,[2,3] Xianfeng Chen,[1,4,5,6] Luqi Yuan,[1,†] and Shanhui Fan[7]

[1]*State Key Laboratory of Advanced Optical Communication Systems and Networks, School of Physics and Astronomy, Shanghai Jiao Tong University, Shanghai 200240, China*

[2]*Interdisciplinary Center for Quantum Information, State Key Laboratory of Modern Optical Instrumentation, and Zhejiang Province Key Laboratory of Quantum Technology and Device, Department of Physics, Zhejiang University, Hangzhou 310027, Zhejiang Province, China*

[3]*CAS Center for Excellence in Topological Quantum Computation, University of Chinese Academy of Sciences, Beijing 100190, China*

[4]*Shanghai Research Center for Quantum Sciences, Shanghai 201315, China*

[5]*Jinan Institute of Quantum Technology, Jinan 250101, China*

[6]*Collaborative Innovation Center of Light Manipulation and Applications, Shandong Normal University, Jinan 250358, China*

[7]*Ginzton Laboratory and Department of Electrical Engineering, Stanford University, Stanford, CA 94305, USA*

[†]yuanluqi@sjtu.edu.cn

**Abstract**

The synthetic dimension, a research topic of both fundamental significance and practical applications, is attracting increasing attention in recent years. In this paper, we propose a theoretical framework to construct arbitrary synthetic dimensions, or *N*-boson synthetic lattices, using multiple bosons on one-dimensional lattices. We show that a one-dimensional lattice hosting *N* indistinguishable bosons can be mapped to a single boson on a *N*-dimensional lattice with high symmetry. Band structure analyses on this *N*-dimensional lattice can then be mathematically performed to predict the existence of exotic eigenstates and the motion of *N*-boson wavepackets. As illustrative examples, we demonstrate the edge states in two-boson Su–Schrieffer–Heeger synthetic lattices without interactions, interface states in two-boson Su–Schrieffer–Heeger synthetic lattices with interactions, and weakly-bound triplon states in three-boson tight-binding synthetic lattices with interactions. The interface states and weakly-bound triplon states have not been thoroughly understood in previous literatures. Our proposed theoretical framework hence provides a novel perspective to explore the multi-boson dynamics on lattices with boson–boson interactions, and opens up a future avenue in the fields of multi-boson manipulation in quantum engineering.

## 1. Introduction

Dimensionality is one of the most important concepts in modern physics. In the condensed-matter physics, systems with different dimensionalities exhibit vast different behaviors. Notable examples include topological insulators with protected surface states [1-4] and superconducting electronic gases confined in quantum structures [5-9], which all exhibit significant dependence of system dimensionality. To explore phenomena unique to high-dimensional physics, the artificial synthesis of extra dimensions in low-dimensional platforms has attracted great interest in recent years [10-14], partly because experimental platforms, including dielectrics [15], plasmons [16,17], atoms [11,18,19], and magnetons [20], feature relatively convenient fabrication and manipulation in low dimensions. By connecting internal degrees of freedom of particles to form apparent artificial lattices [13], one can successfully construct *synthetic dimensions* which, together with spatial dimensions, make the dimensionality of a physical system beyond that of real space. This focus on synthetic dimensions is of fundamental significance in physics, and also holds promising applications in the field of optical communication and quantum information processing [21-25].

Quantum many-body physics, as an important research subject in the quantum physics, receives intensive studies, with its counterparts also greatly explored by quantum simulations in optical platforms [26-28]. As an outstanding example, repulsively bound boson pairs, also called doublons, have been unveiled as a result of Bose–Hubbard Hamiltonian [29]. Following this important discovery, quantum problems of two-particle states with one-dimensional (1D) interactive Hamiltonians, such as the tight-binding Bose–Hubbard model [30], the tight-binding Bose–Hubbard model with a parabolic potential [31], with a small impurity potential [32], and with nonlocal interactions [33-35], Su–Schrieffer–Heeger (SSH) Bose–Hubbard model with nonlocal interactions [36-40], as well as two-particle problems with Bloch oscillation in an external electrical field [41-43], have been investigated with very broad interest in both condensed-matter and optical societies. In particular, when two correlated indistinguishable boson dynamics are under exploration, the *1D–2D mapping* approach [30,32,33,35,39,40,43] has been used to convert the two-boson dynamics on 1D lattices into the single-particle dynamics on two-dimensional (2D) lattices. This mapping method significantly facilitates the theoretical analysis and deepens our understanding of two-boson interacting dynamics in the quantum regime.

Inspired by the works mentioned above, in this paper, we propose to construct arbitrary synthetic dimensions with multiple bosons in a 1D array, i.e., an *N-boson synthetic lattice*, by generalizing the 1D–2D mapping method to a 1D–*N*-dimensional counterpart. After properly choosing the symmetry restriction of the wavefunction, this *N*-dimensional synthetic space with *N* indistinguishable bosons features the periodicity along each of its synthetic dimensions, making it possible to predict the *N*-boson dynamics by utilizing the well-established band structure approach. Related to, but distinct from previous works [44-47], which have constructed extra dimensions beyond the original system by resorting to the photon number on each available lattice site, our theoretical framework harnesses the lattice site number that each individual boson resides on to achieve the dimension augmentation. As for demonstrations, we provide examples exploring multi-particle physics with extended nonlinearity including the two-boson case supporting the topological interface states, as well as the three-boson case predicting the boson blockade effects in three dimensions. Our work hence points out a fundamentally different perspective for studying many-body dynamics with nonlinearity from a physical picture with synthetic dimensions, which also shows potential applications in quantum simulations and quantum information processing with synthetic dimensions in optical systems.

The remaining parts of this paper are organized as follows. Section 2 describes the fundamental theoretical framework of the *N*-boson synthetic lattice that will be adopted throughout this paper. Sections 3, 4 and 5 give examples addressing the richness of various systems where our theoretical framework is applicable. Section 6 discusses some possible experimental implementations followed by a brief conclusion.

## 2. Theoretical framework

We start to establish our theoretical framework by examining two indistinguishable bosons on an infinite 1D tight-binding lattice, and then move on to the *N*-boson case. The two-boson case has been addressed in several previous researches as the 1D–2D mapping approach [30,32,33,35,39,40,43]. Here we first re-iterate this approach in a more explicit and systematic way. The Hamiltonian of the system is:

$$H = g \sum_{k=-\infty}^{+\infty} \left( b_k^\dagger b_{k+1} + b_{k+1}^\dagger b_k \right), \tag{1}$$

where *g*, a real number, is the coupling strength between neighboring sites, and $b_k$ ($b_k^\dagger$) is the annihilation (creation) operator of bosons on the *k*-th lattice site as shown in Fig. 1(a). We take $\hbar = 1$ throughout the paper for simplicity. In Schrödinger's picture, the two-boson state is:

$$|\phi(t)\rangle = \frac{1}{\sqrt{2}} \sum_{m=-\infty}^{+\infty} \sum_{n=-\infty}^{+\infty} v_{mn}(t) b_m^\dagger b_n^\dagger |0\rangle, \tag{2}$$

where $|0\rangle$ is the vacuum state. By substituting Eqs. (1) and (2) into the Schrödinger equation $i|\dot\phi\rangle = H|\phi\rangle$, the differential equation satisfied by $v_{mn}$ is derived as

$$i(\dot v_{mn} + \dot v_{nm}) = g\left( v_{m-1,n} + v_{m+1,n} + v_{m,n-1} + v_{m,n+1} + v_{n-1,m} + v_{n+1,m} + v_{n,m-1} + v_{n,m+1} \right). \tag{3}$$

Here, the conventionally standard restriction on $v_{mn}$ is to follow the *exchange symmetry*, i.e., we let $v_{mn} = v_{nm}$. With this restriction,

$$\langle 0|b_m b_n|\phi\rangle = \frac{1}{\sqrt{2}} \sum_{m'=-\infty}^{+\infty} \sum_{n'=-\infty}^{+\infty} v_{m'n'} \langle 0|b_m b_n b_{m'}^\dagger b_{n'}^\dagger|0\rangle = \frac{v_{mn} + v_{nm}}{\sqrt{2}} = \sqrt{2} v_{mn}, \tag{4}$$

and the square of the modulus of $v_{mn}$ can then be identified as the second-order correlation function of the two-boson state:

$$|v_{mn}|^2 = \frac{1}{2}\langle\phi|b_n^\dagger b_m^\dagger|0\rangle\langle 0|b_m b_n|\phi\rangle = \frac{1}{2}\langle\phi|b_n^\dagger b_m^\dagger b_m b_n|\phi\rangle. \tag{5}$$

Also, the probability amplitude that one boson is observed on the *m*-th lattice site and the other on the *n*-th site ($m \leq n$) is

$$u_{mn} = \begin{cases} \langle 1_m 1_n|\phi\rangle = \sqrt{2}v_{mn} = \sqrt{2}v_{nm} & \text{if } m < n \\ \langle 2_m|\phi\rangle = v_{mm} & \text{if } m = n \end{cases}. \tag{6}$$

Since we have required that $v_{mn} = v_{nm}$, Eq. (3) simplifies to

$$i\dot{v}_{mn} = g\left(v_{m-1,n} + v_{m+1,n} + v_{m,n-1} + v_{m,n+1}\right), \tag{7}$$

which shows that the dynamics of $v_{mn}$ is mathematically equivalent to the single-particle dynamics on a 2D tight-binding lattice. This lattice, referred to as a two-boson synthetic lattice, is illustrated in Fig. 1(b), and the process of constructing such two-boson synthetic lattice can be denoted as 1D–2D mapping.

The procedure for the construction of such synthetic lattice can be generalized for *N*-boson states. We consider the tight-binding Hamiltonian of Eq. (1) operating on an *N*-boson state

$$|\phi(t)\rangle = \frac{1}{\sqrt{N!}} \sum_{\lambda_1=-\infty}^{+\infty} \cdots \sum_{\lambda_N=-\infty}^{+\infty} \left[v_{\lambda_1,\cdots,\lambda_N}(t) b_{\lambda_1}^\dagger \cdots b_{\lambda_N}^\dagger\right]|0\rangle. \tag{8}$$

From the Schrödinger equation, we can calculate that

$$i \sum_{\text{permu}} \dot{v}_{\lambda_1,\cdots,\lambda_N} = g \sum_{\text{permu}} \sum_{l=1}^{N} \left(v_{\lambda_1,\cdots,\lambda_{l-1},\lambda_l+1,\lambda_{l+1},\cdots,\lambda_N} + v_{\lambda_1,\cdots,\lambda_{l-1},\lambda_l-1,\lambda_{l+1},\cdots,\lambda_N}\right), \tag{9}$$

where $\sum_{\text{permu}}(\square)$ denotes the permutation sum by re-arranging the sequence $\lambda_1,\cdots,\lambda_N$. Similarly here, the wavefunction is additionally restricted by the exchange symmetry that $v_{\lambda_1,\cdots,\lambda_N}$ remains constant after exchanging any two subscripts $\lambda_i$ and $\lambda_j$. In this case,

$$\langle 0|b_{\lambda_1}\cdots b_{\lambda_N}|\phi\rangle = \frac{\sum_{\text{permu}} v_{\lambda_1,\cdots,\lambda_N}}{\sqrt{N!}} = \sqrt{N!}\, v_{\lambda_1,\cdots,\lambda_N}, \tag{10}$$

and the square of the modulus of $v_{\lambda_1,\cdots,\lambda_N}$ is the *N*-th-order correlation function:

$$|v_{\lambda_1,\cdots,\lambda_N}|^2 = \frac{1}{N!}\langle\phi|b_{\lambda_N}^\dagger \cdots b_{\lambda_1}^\dagger b_{\lambda_1} \cdots b_{\lambda_N}|\phi\rangle. \tag{11}$$

Using $v_{\lambda_1,\cdots,\lambda_N}$, the probability amplitude that $\xi_i$ bosons are observed on the $\mu_i$-th lattice site ($\xi_i > 0$, $\sum_i \xi_i = N$, and $\mu_i < \mu_j$ if $i < j$) can be expressed as

$$u_{\underbrace{\mu_1,\mu_1,\cdots,\mu_1}_{\xi_1},\cdots,\underbrace{\mu_i,\mu_i,\cdots,\mu_i}_{\xi_i},\cdots} = \langle \xi_{1\mu_1} \cdots \xi_{i\mu_i} \cdots | \phi \rangle = \sqrt{\frac{N!}{\prod_i \xi_i!}} v_{\underbrace{\mu_1,\mu_1,\cdots,\mu_1}_{\xi_1},\cdots,\underbrace{\mu_i,\mu_i,\cdots,\mu_i}_{\xi_i},\cdots}. \quad (12)$$

Because of the exchange symmetry, Eq. (9) can finally be simplified to

$$i\dot{v}_{\lambda_1,\lambda_2,\cdots,\lambda_N} = g\left[\left(v_{\lambda_1+1,\lambda_2,\cdots,\lambda_N} + v_{\lambda_1,\lambda_2+1,\cdots,\lambda_N} + \cdots + v_{\lambda_1,\lambda_2,\cdots,\lambda_N+1}\right) + \left(v_{\lambda_1-1,\lambda_2,\cdots,\lambda_N} + v_{\lambda_1,\lambda_2-1,\cdots,\lambda_N} + \cdots + v_{\lambda_1,\lambda_2,\cdots,\lambda_N-1}\right)\right]. \quad (13)$$

Eq. (13) is mathematically identical to the coupled-mode equations [48] for the wave evolution in an isotropic lattice defined in N dimensions, thus giving the N-boson synthetic lattice with $v_{\lambda_1,\cdots,\lambda_N}$ being the lattice sites.

In deriving Eq. (13), we assume that the wavefunction $v_{\lambda_1,\cdots,\lambda_N}$ satisfies exchange symmetry. For the same quantum state of Eq. (8), other restrictions of the wavefunction $v_{\lambda_1,\cdots,\lambda_N}$ are also possible. For example, we may also choose $v_{\lambda_1,\cdots,\lambda_N}$ to be equal to zero except in the region $\lambda_1 \leq \lambda_2 \leq \cdots \leq \lambda_N$. With this alternative choice, however, Eq. (13) no longer holds, and $v_{\lambda_1,\cdots,\lambda_N}$'s appear to be connected to their neighboring sites with non-uniform coupling strengths (see Supplemental Notes 1 [49]). In fact, the uniformity of the lattice as described by Eq. (13) is closely related to imposing the exchange symmetry on the wavefunction.

The uniformity of the lattice as described in Eq. (13) allows the definition of a band structure in the synthetic space. The steady state solution of Eq. (13) corresponds to the following eigenvalue problem:

$$\varepsilon_{\text{tight-binding}}^{N\text{-dimension}} v_{\lambda_1,\lambda_2,\cdots,\lambda_N} = g\left[\left(v_{\lambda_1+1,\lambda_2,\cdots,\lambda_N} + v_{\lambda_1,\lambda_2+1,\cdots,\lambda_N} + \cdots + v_{\lambda_1,\lambda_2,\cdots,\lambda_N+1}\right) + \left(v_{\lambda_1-1,\lambda_2,\cdots,\lambda_N} + v_{\lambda_1,\lambda_2-1,\cdots,\lambda_N} + \cdots + v_{\lambda_1,\lambda_2,\cdots,\lambda_N-1}\right)\right]. \quad (14)$$

Since the coupling strengths are uniform in this synthetic lattice along all of the N dimensions ($\lambda_1,\cdots,\lambda_N$), we can define an N-boson band structure analytically:

$$\varepsilon_{\text{tight-binding}}^{N\text{-dimension}}(\mathbf{k}) = 2g\sum_{i=1}^{N}\cos k_i, \quad (15)$$

where $\mathbf{k} = [k_1,\cdots,k_N]^{\text{T}}$ and $k_i$ is the wavevector along the $\lambda_i$ direction in Eq. (14). The group velocity of the particle wavepacket in the synthetic lattice associated with a specific N-boson state can be calculated using the band structure by computing the derivative of the energy spectrum with respect to the wavevectors. One notices that the band structure for a 1D tight-binding lattice is $\varepsilon_{\text{tight-binding}}^{1D}(k) = 2g\cos k$ [50] so that $\varepsilon_{\text{tight-binding}}^{N\text{-dimension}}(\mathbf{k}) = \sum_{i=1}^{N}\varepsilon_{\text{tight-binding}}^{1D}(k_i)$. This result is consistent with the physical intuition that, without the boson–boson interaction, $N$ bosons evolve independently so the $N$-boson dynamics can be understood from the single-boson dynamics. On the other hand, when the boson–boson interaction is involved, the $N$-boson dynamics can no longer be simply derived from the one-boson dynamics. In later sections of this paper, we shall study examples with nonlinear boson–boson interactions, where the $N$-boson synthetic lattice shows great convenience in predicting some intriguing phenomena.

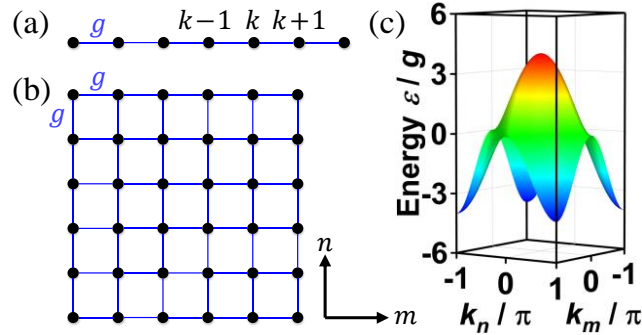

**Fig. 1. The two-boson tight-binding synthetic lattice without interactions.** (a) A 1D tight-binding lattice with coupling strength $g$. (b) The two-boson synthetic lattice with the exchange symmetry restriction on the wavefunction $v_{mn}(t)$. (c) 2D band structure of the lattice in (b) in the first Brillouin zone.

To confirm the predictions made by the $N$-boson synthetic lattice and the corresponding band structure analysis, we can also perform numerical simulations by solving Eq. (13) in the presence of external excitation. Based on the input–output formalism [51], when $\xi_i$ bosons are excited on the $\mu_i$-th lattice site ($\xi_i > 0$, $\sum_i \xi_i = N$, and $\mu_i < \mu_j$ if $i < j$), a source term $\eta(t)$ should be added to the right side of Eq. (13) if $\lambda_1, \cdots, \lambda_N$ is a permutation of $\underbrace{\mu_1, \mu_1, \cdots, \mu_1}_{\xi_1}, \cdots, \underbrace{\mu_i, \mu_i, \cdots, \mu_i}_{\xi_i}, \cdots$. Here $\eta(t)$, the temporal profile of the boson excitation, takes the form of a modulated Gaussian envelope

$$\eta(t) = \eta_0 \exp\left(-\frac{(t-t_0)^2}{\tau^2}\right) \exp(-i\Delta\varepsilon t),\ t \geq 0, \qquad (16)$$

where $\eta_0$ is the normalization coefficient, $t_0$ and $\tau$ are the center and the width of the Gaussian pulse respectively, and $\Delta\varepsilon$ is the excitation energy. Finally, the multi-boson excitation probabilities can be calculated according to Eq. (12), and the average boson number on the $k$-th lattice site of the hosting 1D tight-binding chain (Fig. 1(a)) can be determined by

$$N_k(t) = \langle \phi(t) | b_k^\dagger b_k | \phi(t) \rangle. \tag{17}$$

Some typical simulation results of the two-boson dynamics with Hamiltonian Eq. (1) are displayed in Fig. S2 [49].

Before ending this section, we note that, for illustration purposes, the synthetic lattice described in this section is rather simple with only real-valued, nearest-neighbor coupling and no interaction. However, more complex behaviors including non-trivial topological and interacting features can be created in the *N*-boson synthetic lattice based on the same theoretical framework as introduced here. To highlight these more complex features, in the following, we discuss two bosons on an SSH lattice without/with interactions (Sections 3 and 4), and two/three bosons on a tight-binding extended Bose–Hubbard lattice (Supplemental Notes 2 [49] and Section 5).

## 3. Two bosons on an SSH lattice without interactions

As the first illustration of interesting topological behaviors in the *N*-boson synthetic lattice, here we examine the two-boson dynamics on a 1D SSH lattice. The 1D SSH model is perhaps one of the simplest models that exhibit a nontrivial energy band topology [52], and has been widely studied [53-58]. Here we show that dynamics of two non-interacting bosons on a 1D SSH lattice can be mapped to a 2D SSH model. A 1D SSH lattice, as shown in Fig. 2(a), features alternating coupling strengths of $g_1$ and $g_2$, and consequently there are two types of lattice sites on the chain which are denoted as $C_k$ and $D_k$ in the *k*-th unit cell, respectively. $g_1$ is referred to as intracell coupling and $g_2$ as intercell coupling. The corresponding Hamiltonian and the two-boson state under consideration are expressed by

$$H = g_1 \sum_{k=-\infty}^{+\infty} \left( c_k^\dagger d_k + d_k^\dagger c_k \right) + g_2 \sum_{k=-\infty}^{+\infty} \left( c_{k+1}^\dagger d_k + d_k^\dagger c_{k+1} \right), \tag{18}$$

$$|\phi(t)\rangle = \frac{1}{\sqrt{2}} \sum_{m=-\infty}^{+\infty} \sum_{n=-\infty}^{+\infty} \left[ v_{2m,2n}(t) c_m^\dagger c_n^\dagger + v_{2m,2n+1}(t) c_m^\dagger d_n^\dagger + v_{2m+1,2n}(t) d_m^\dagger c_n^\dagger + v_{2m+1,2n+1}(t) d_m^\dagger d_n^\dagger \right] |0\rangle, \tag{19}$$

where $c_k$ and $d_k$ ($c_k^\dagger$ and $d_k^\dagger$) are annihilation (creation) operators of bosons on the $C_k$ and $D_k$ sites, respectively. Substituting Eqs. (18) and (19) into the Schrödinger equation $i|\dot\phi\rangle = H|\phi\rangle$, we obtain the differential equations:

$$\begin{cases} i(\dot v_{2m,2n} + \dot v_{2n,2m}) = g_1(v_{2m,2n+1} + v_{2n,2m+1} + v_{2m+1,2n} + v_{2n+1,2m}) + g_2(v_{2m,2n-1} + v_{2n,2m-1} + v_{2m-1,2n} + v_{2n-1,2m}) \\ i(\dot v_{2m+1,2n+1} + \dot v_{2n+1,2m+1}) = g_1(v_{2m,2n+1} + v_{2n,2m+1} + v_{2m+1,2n} + v_{2n+1,2m}) + g_2(v_{2m+2,2n+1} + v_{2n+2,2m+1} + v_{2m+1,2n+2} + v_{2n+1,2m+2}) \\ i(\dot v_{2m,2n+1} + \dot v_{2n+1,2m}) = g_1(v_{2m,2n} + v_{2n,2m} + v_{2m+1,2n+1} + v_{2n+1,2m+1}) + g_2(v_{2m,2n+2} + v_{2n+2,2m} + v_{2m-1,2n+1} + v_{2n+1,2m-1}) \end{cases} . \tag{20}$$

Now we apply the exchange symmetry restriction of the wavefunction ($v_{mn} = v_{nm}$ as discussed in Section 2), and Eq. (20) can then be transformed by into

$$\begin{cases} i\dot{v}_{2m,2n} = g_1\left(v_{2m,2n+1} + v_{2m+1,2n}\right) + g_2\left(v_{2m,2n-1} + v_{2m-1,2n}\right) \\ i\dot{v}_{2m,2n+1} = g_1\left(v_{2m,2n} + v_{2m+1,2n+1}\right) + g_2\left(v_{2m,2n+2} + v_{2m-1,2n+1}\right) \\ i\dot{v}_{2m+1,2n} = g_1\left(v_{2m,2n} + v_{2m+1,2n+1}\right) + g_2\left(v_{2m+2,2n} + v_{2m+1,2n-1}\right) \\ i\dot{v}_{2m+1,2n+1} = g_1\left(v_{2m,2n+1} + v_{2m+1,2n}\right) + g_2\left(v_{2m+2,2n+1} + v_{2m+1,2n+2}\right) \end{cases}. \quad (21)$$

Again, this exchange symmetry is not a physically necessary restriction on the wavefunction, yet it is mathematically essential in obtaining Eq. (21). Eq. (21) gives the two-boson synthetic lattice as illustrated in Fig. 2(b), which corresponds to a standard 2D SSH model [59]. Mathematically, this lattice features translational symmetry along both $m$ and $n$ axes with period 2, and its band structure can be analytically obtained as:

$$\varepsilon_{\text{SSH}}^{\text{2D}}(k_m, k_n) = \pm\sqrt{g_1^2 + g_2^2 + 2g_1g_2\cos k_m} \pm \sqrt{g_1^2 + g_2^2 + 2g_1g_2\cos k_n}, \quad (22)$$

as plotted in Fig. 2(c) [60]. Four different bands are observed as one unit cell in the two-boson synthetic lattice contains four lattice sites: $v_{2m,2n}$, $v_{2m,2n+1}$, $v_{2m+1,2n}$ and $v_{2m+1,2n+1}$. The band structure of Eq. (22) can also be expressed as

$$\varepsilon_{\text{SSH}}^{\text{2D}}(k_m, k_n) = \varepsilon_{\text{SSH}}^{\text{1D}}(k_m) + \varepsilon_{\text{SSH}}^{\text{1D}}(k_n), \quad (23)$$

where $\varepsilon_{\text{SSH}}^{\text{1D}}(k) = \pm\sqrt{g_1^2 + g_2^2 + 2g_1g_2\cos k}$, because no boson–boson interaction is considered in Hamiltonian Eq. (18) [61].

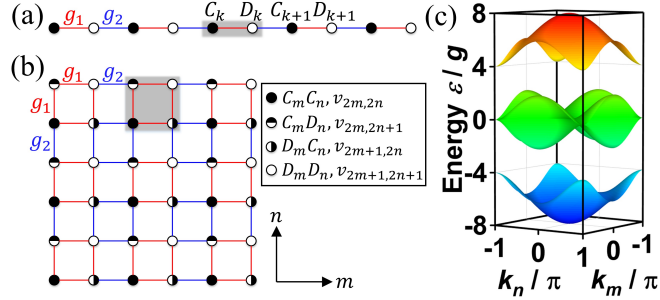

**Fig. 2. The two-boson SSH synthetic lattice without interactions.** (a) A 1D SSH lattice, where black spots are sites $C_k$'s, white spot are sites $D_k$'s, red lines are intracell coupling $g_1$ and blue lines are intercell coupling $g_2$. One unit cell contains two sites as shaded in grey. (b) The two-boson synthetic lattice with the exchange symmetry restriction on the wavefunction $v_{2m,2n}(t)$, $v_{2m,2n+1}(t)$, $v_{2m+1,2n}(t)$ and $v_{2m+1,2n+1}(t)$. One unit cell contains four sites as shaded in grey. (c) 2D band structure of the lattice in (b) in the first Brillouin zone with coupling strengths $g_1 = 3g$, $g_2 = g$.

A 2D SSH lattice exhibits topological features since edge modes exist in the nontrivial phase ($g_1 < g_2$) because of nonzero Berry connection and Zak phase [60]. Here we examine an SSH stripe which is infinite along the $m$ axis but has finite width along the $n$ axis ($-15 \leq n \leq 15$). (Strictly speaking, the trivial/nontrivial phase transition point is not exactly at $g_1 = g_2$, depending on the stripe width, but takes $g_1 = g_2$ in the limit of infinite stripe width [60].) One notes that such an SSH stripe is not physically implementable by our platform of two bosons on a 1D lattice, since the wavefunction exchange symmetry requires that the lattice should have reflection symmetry along $m = n$ axis; yet this SSH stripe is still a useful mathematical construct for analyzing edge modes. In Figs. 3(a) and 3(b) the projected band structures are plotted in the topologically trivial phase ($g_1 = 3g$, $g_2 = g$) and nontrivial phase ($g_1 = g$, $g_2 = 3g$), respectively. The bulk modes (drawn in blue) can be viewed as Fig. 2(c) after the band structure is projected onto the $k_m$ axis. In the nontrivial phase, edge modes (red lines) are clearly shown in the band gaps. Eigenstate distribution analysis further confirms that these isolated modes are edge modes, since the corresponding eigenstates are localized at lattice sites whose indices $n$'s are $\pm 15$, i.e. on the edges of the SSH stripe, as indicated by black arrows in Figs. 3(c)–3(f).

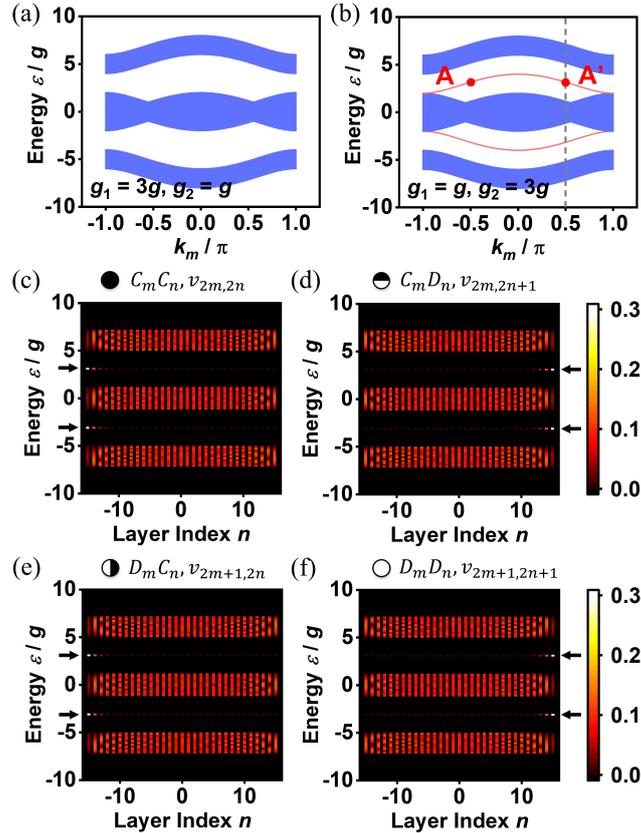

**Fig. 3. Projected band structures and eigenstate distributions of an SSH stripe.** (a)–(b) Projected band structures of an SSH stripe with coupling strengths $(g_1, g_2) = (3g, g)$ or $(g, 3g)$. Edge states are denoted by red lines in (b). (c)–(f) Eigenstate distributions of (c) $C_m C_n$-type, (d) $C_m D_n$-type, (e) $D_m C_n$-type and (f) $D_m D_n$-type lattice sites corresponding to the band structures (b) with $k_m = 0.5\pi$

(grey dash in (b)). Edge modes are indicated by black arrows. $-\infty < m < +\infty$, $-15 \leq n \leq 15$ in this figure.

Although the band structure analysis shown in Fig. 3 is based on an infinite SSH stripe, the edge modes still preserve in a finite two-boson synthetic lattice (i.e. a square-shape 2D SSH panel as in Fig. 2(b)), which we demonstrate with numerical simulations in Fig. 4. In simulations, we consider a 1D SSH lattice with 62 lattice sites (31 unit cells), i.e., $-15 \leq k \leq 15$ in Fig. 2(a). It gives a two-boson synthetic lattice with $62 \times 62$ sites in two dimensions. For the convenience of plotting in Fig. 4, lattice site $C_k$ in Fig. 2(a) is labeled as the 2k-th site (i.e., $C_k$ sites have indices $2k = -30, -28, \cdots, 30$) and $D_k$ as the (2k + 1)-th site (i.e., $D_k$ sites have indices $2k+1 = -29, -27, \cdots, 31$). We numerically solve Eq. (21) together with the input–output formalism following the procedures described in Section 2. The probability amplitude that two bosons are respectively observed on lattice sites $C_m$ and $C_n$ ($D_m$ and $D_n$, $C_m$ and $D_n$) in Fig. 2(a) is

$$\begin{cases} (C_m, C_n): \lambda_{mn}(t) = \sqrt{2-\delta_{mn}} v_{2m,2n} & (-15 \leq m \leq n \leq 15) \\ (D_m, D_n): \mu_{mn}(t) = \sqrt{2-\delta_{mn}} v_{2m+1,2n+1} & (-15 \leq m \leq n \leq 15), \\ (C_m, D_n): \xi_{mn}(t) = \sqrt{2} v_{2m,2n+1} & (-15 \leq m, n \leq 15) \end{cases} \quad (24)$$

where $\delta_{mn}$ is the Kronecker delta function, and the average boson number on each site of the 1D SSH lattice can be calculated as:

$$\text{Site } C_k: N_{2k}(t) = \sum_{j=-15}^{k-1} |\lambda_{jk}(t)|^2 + 2|\lambda_{kk}(t)|^2 + \sum_{j=k+1}^{15} |\lambda_{kj}(t)|^2 + \sum_{j=-15}^{15} |\xi_{kj}(t)|^2, \quad (25)$$

$$\text{Site } D_k: N_{2k+1}(t) = \sum_{j=-15}^{k-1} |\mu_{jk}(t)|^2 + 2|\mu_{kk}(t)|^2 + \sum_{j=k+1}^{15} |\mu_{kj}(t)|^2 + \sum_{j=-15}^{15} |\xi_{jk}(t)|^2. \quad (26)$$

We plot evolutions of the average boson numbers on the 1D SSH lattice in Figs. 4(a), 4(c), 4(e), while plotting the corresponding distributions of the excitation probabilities on the two-boson synthetic lattice in Figs. 4(b), 4(d), 4(f), respectively. In Figs. 4(a) and 4(b), we perform the simulation in the nontrivial phase and apply the external source in Eq. (16) with $\Delta\varepsilon = 3.16g$ to excite two bosons on lattice sites $C_0$ and $D_{15}$ (i.e., the 0-th and 31-st sites in the 1D lattice), which corresponds to the excitation at the site (0, 31) in the two-boson synthetic lattice in Fig. 4(b). One clearly sees the edge modes in the two-boson synthetic lattice in Fig. 4(b), which features bi-directional propagation because the corresponding mode A (with a positive slope) and A' (with a negative slope) in Fig. 3(b) are both excited by the source. The corresponding evolution of the average boson numbers in the 1D lattice in Fig. 4(a) exhibits the same feature, where one boson excited at the boundary is localized at $D_{15}$, while the other boson excited at $C_0$ propagates along both directions in one dimension. The band analysis based on the two-boson synthetic lattice is therefore useful for us to understand the evolution of the boson number distribution along the 1D lattice. As a comparison, if we excite the bulk site (0, 0) in the nontrivial phase (Figs. 4(c) and 4(d)), or excite the edge site (0, 31) in the trivial phase (Figs. 4(e) and 4(f)) in simulations, no edge modes are observed.

We shall clarify that, as mentioned in Eq. (23), without boson–boson interaction, the two-boson dynamics on a 1D SSH lattice can be readily inferred from the single-boson dynamics on a 1D SSH lattice. For example, in Fig. 4(a), the two-boson edge mode is in fact one boson localized at the boundary of the original 1D SSH chain and the other propagating inside the 1D chain. In another example of Fig. S6, our two-boson synthetic lattice shows that a two-boson corner mode is simply the combination of two single-boson boundary modes, where both bosons are localized on either terminal of the 1D SSH chain in the nontrivial phase [49].

In this section, we utilize a simple example to show that a perspective from a synthetic lattice helps one to understand the $N$-boson dynamics in a 1D lattice. This capability will be much more important when the interaction between bosons is introduced, which we will show in the next two sections.

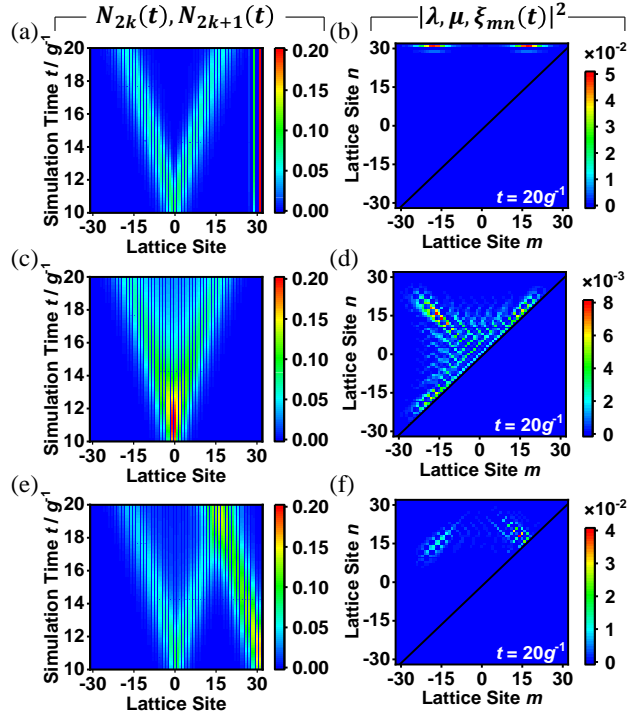

**Fig. 4. Simulation results of edge modes in the two-boson SSH synthetic lattice without interactions.** (a)–(b) Excite two bosons on sites $C_0$ and $D_{15}$ respectively with $g_1 = g$, $g_2 = 3g$ and $\Delta\varepsilon = 3.16g$. (c)–(d) Excite two bosons on site $C_0$ simultaneously with $g_1 = g$, $g_2 = 3g$ and $\Delta\varepsilon = 0$. (e)–(f) Excite two bosons on sites $C_0$ and $D_{15}$ respectively with $g_1 = 3g$, $g_2 = g$ and $\Delta\varepsilon = 0$. $t_0 = 10g^{-1}$ and $\tau^2 = 10g^{-2}$ in this figure.

## 4. Two bosons on an SSH lattice with interactions

An interacting quantum system certainly show a far richer physical behavior as compared with a non-interacting systems. The study of the *N*-boson synthetic lattice provides a unique perspective to explore multi-boson problems with interactions. As an illustration, we consider two bosons on the same 1D SSH lattice as that in Section 3, but with boson–boson interactions of the extended Bose–Hubbard type. The corresponding Hamiltonian is described by

$$H = \sum_{k=-\infty}^{+\infty} \left( g_1 c_k^\dagger d_k + g_2 c_{k+1}^\dagger d_k + \text{H.c.} \right) + \frac{U}{2} \sum_{k=-\infty}^{+\infty} \sum_{s=-R}^{+R} \left( c_{k+s}^\dagger c_k^\dagger c_k c_{k+s} + d_{k+s}^\dagger d_k^\dagger d_k d_{k+s} + 2 c_{k+s}^\dagger d_k^\dagger d_k c_{k+s} \right) \quad , \tag{27}$$

where $U$ is the interaction strength and $R$ is the interaction range. Different from previous works [36-39], here we show the capability of our method to understand the multi-boson problem with long-range nonlocal interactions, and predict the existence of novel topological states that have not been explored before. For this purpose, we take the same two-boson state in Eq. (19) and apply the Schrödinger equation with the Hamiltonian in Eq. (27) (assuming $R \geq 2$), and obtain

$$\begin{cases} i\left(\dot{v}_{2m,2n} + \dot{v}_{2n,2m}\right) = g_1 \left(v_{2m,2n+1} + v_{2n,2m+1} + v_{2m+1,2n} + v_{2n+1,2m}\right) + g_2 \left(v_{2m,2n-1} + v_{2n,2m-1} + v_{2m-1,2n} + v_{2n-1,2m}\right) + U\delta_{|m-n|\leq R}\left(v_{2m,2n} + v_{2n,2m}\right) \\ i\left(\dot{v}_{2m+1,2n+1} + \dot{v}_{2n+1,2m+1}\right) = g_1 \left(v_{2m,2n+1} + v_{2n,2m+1} + v_{2m+1,2n} + v_{2n+1,2m}\right) + g_2 \left(v_{2m+2,2n+1} + v_{2n+2,2m+1} + v_{2m+1,2n+2} + v_{2n+1,2m+2}\right) + U\delta_{|m-n|\leq R}\left(v_{2m+1,2n+1} + v_{2n+1,2m+1}\right) \\ i\left(\dot{v}_{2m,2n+1} + \dot{v}_{2n+1,2m}\right) = g_1 \left(v_{2m,2n} + v_{2n,2m} + v_{2m+1,2n+1} + v_{2n+1,2m+1}\right) + g_2 \left(v_{2m,2n+2} + v_{2n+2,2m} + v_{2m-1,2n+1} + v_{2n+1,2m-1}\right) + U\delta_{|m-n|\leq R}\left(v_{2m,2n+1} + v_{2n+1,2m}\right) \end{cases}, \tag{28}$$

where $\delta_{|m-n|\leq R} = 1$ if $|m-n| \leq R$ and $= 0$ otherwise. Imposing the exchange symmetry on the wavefunction ($v_{mn} = v_{nm}$), we can re-write Eq. (28) as

$$\begin{cases} i\dot{v}_{2m,2n} = g_1 \left(v_{2m,2n+1} + v_{2m+1,2n}\right) + g_2 \left(v_{2m,2n-1} + v_{2m-1,2n}\right) + U\delta_{|m-n|\leq R} v_{2m,2n} \\ i\dot{v}_{2m,2n+1} = g_1 \left(v_{2m,2n} + v_{2m+1,2n+1}\right) + g_2 \left(v_{2m,2n+2} + v_{2m-1,2n+1}\right) + U\delta_{|m-n|\leq R} v_{2m,2n+1} \\ i\dot{v}_{2m+1,2n} = g_1 \left(v_{2m,2n} + v_{2m+1,2n+1}\right) + g_2 \left(v_{2m+2,2n} + v_{2m+1,2n-1}\right) + U\delta_{|m-n|\leq R} v_{2m+1,2n} \\ i\dot{v}_{2m+1,2n+1} = g_1 \left(v_{2m,2n+1} + v_{2m+1,2n}\right) + g_2 \left(v_{2m+2,2n+1} + v_{2m+1,2n+2}\right) + U\delta_{|m-n|\leq R} v_{2m+1,2n+1} \end{cases}. \tag{29}$$

Eq. (29) therefore again can be mapped to a synthetic lattice in two dimensions as shown in Fig. 5(a). Here *U* appears as the onsite potential on the diagonal with a width of 2*R* + 1 in this two-boson synthetic lattice, as a result of the nonlinear interaction terms in Eq. (27), as indicated by the green-colored lattice sites in Fig. 5(a). These lattice sites with nonzero onsite potentials compose the 'nonlinear lattice region', while the sites without onsite potentials compose the 'linear lattice region'. Fig. 5(a) features translational symmetry along the *j* axis ($\hat{\mathbf{j}} = (\hat{\mathbf{m}} + \hat{\mathbf{n}})/\sqrt{2}$), and the corresponding wavevector $k_j$ is in fact the quasi-momentum of the motion of the center of mass of these two bosons ($k_j = k_m + k_n$) [29].

For a stripe with finite width along the *l* axis ($\hat{\mathbf{l}} = (\hat{\mathbf{n}} - \hat{\mathbf{m}})/\sqrt{2}$, perpendicular to the *j* axis) and infinite length along the *j* axis, we perform Fourier transform along the *j* axis and then plot the projected band structure $\varepsilon(k_j)$ with different combinations of parameters $g_1$, $g_2$, and $U$. When $U=0$, $g_1 = 3g$, and $g_2 = g$, the projected band structure can be viewed as the 2D SSH band structure after projection along $\Gamma$–M direction in the first Brillouin zone and then band folding (see Figs. S4(c) and S4(d) in the Supplemental Material [49]). Then we plot the projected band structure for the case $U = 2g$, $R = 6$, $g_1 = 3g$, $g_2 = g$ (trivial phase) in Fig. 5(b). One can see that, compared to the plot in Fig. S4(d) [49], part of the bands inside the quasi-continuum are lifted by $2g$, which is equal to the onsite potential $U$. This observation is consistent with the results in Supplemental Notes 2 [49], i.e., the lifted bands with higher energies correspond to two-boson bound states (doublons) in the nonlinear lattice region around $l = 0$, and the unlifted bands with lower energies correspond to free-particle scattering states in the linear region. When $U = 2g$, $R = 6$, $g_1 = g$, $g_2 = 3g$ (nontrivial phase), we plot the band structure in Fig. 5(c) and can see topological effects. The edge mode in the upper band gap is identified and indicated in red, yet the edge mode in the lower band gap is overlapped by lifted bulk bands (i.e. the edge mode 'collapses' into scattering bulk modes). Moreover, there are other isolated bands found around $k_j = 0$ and $\varepsilon = g$. We plot the eigenstate distributions of $C_m D_n$ sites with $k_j = 0$ for the trivial and nontrivial phases in Figs. 5(d) and 5(e), respectively. (Other eigenstate distributions of $C_m C_n$, $D_m C_n$ and $D_m D_n$ sites can be found in Fig. S7 [49].) One can see that, for the nontrivial case, there exists a pair of localized states near $\varepsilon = g$ at the interface between linear and nonlinear lattice regions. These states are the *interface modes*. Here the interaction range $R$ must be large enough so that the isolated bands of interface modes exist and the modes at the two interfaces do not hybridize.

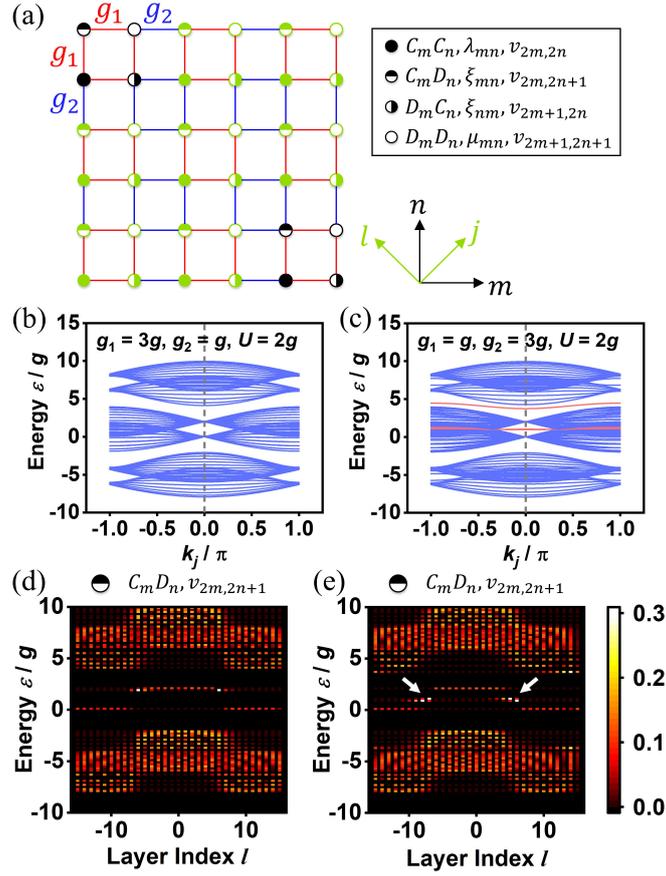

**Fig. 5. The two-boson SSH synthetic lattice with interactions.** (a) The two-boson synthetic lattice with the exchange symmetry restriction on the wavefunction, where green spots indicate nonzero onsite potentials. (b)–(c) Projected band structures of the lattice in (a) with $U = 2g$, $(g_1, g_2) = (3g, g)$ or $(g, 3g)$. (d)–(e) Eigenstate distributions of $C_m D_n$-type lattice sites corresponding to the band structures in (b) and (c) at $k_j = 0$ (grey dashes in (b) and (c)), in topologically (d) trivial and (e) nontrivial phases. Interface modes are indicated by white arrows. $R = 6$ and $-15 \leq l \leq 15$ in (b)–(e).

To see details of the band structure in Fig. 5(c), we focus on the region around $k_j = 0$ and $\varepsilon = g$ in Fig. 6(a). One can see that these interface modes are composed of two groups of degenerate bands (bands 1, 2 and bands 3, 4) with slightly different energies. Figs. 6(b)–6(e) plot the eigenstates distributions of the corresponding four bands, respectively, in a two-boson synthetic lattice with parameters $-2 \leq m, n \leq 3$ and $R = 2$ for the illustration purpose. The lower-energy group of bands corresponds to the modes on the linear side of the interface (bands 1, 2, Figs. 6(b) and 6(c)), while the higher-energy group corresponds to the modes on the nonlinear side of the interface (bands 3, 4, Figs. 6(d) and 6(e)). It is shown that, of the two degenerate bands in each group, one band is symmetric with respect to the line $m = n$ (Figs. 6(c) and 6(e)) and the other one is anti-symmetric (Figs. 6(b) and 6(d)). However, only the symmetric modes can be physically excited because of our exchange symmetry restriction of the wavefunction. The pair of anti-symmetric interface modes (bands 1, 3) is an artifact produced from the 1D–2D mapping method, and does not physically exist.

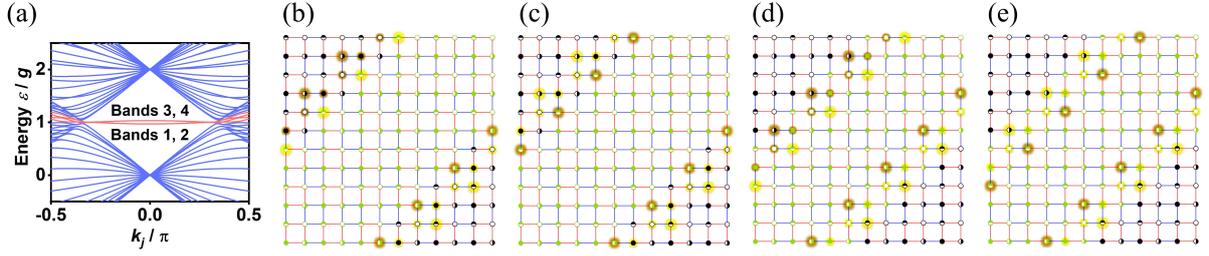

**Fig. 6. Eigenstate distributions of interface modes in the two-boson SSH synthetic lattice with interactions.** (a) Zoom-in view of the projected band structure Fig. 5(c), with $g_1 = g$, $g_2 = 3g$, $U = 2g$, $R = 6$ and $-15 \leq l \leq 15$. (b)–(e) Eigenstate distributions of (b) band 1 (anti-symmetric band with lower energy), (c) band 2 (symmetric band with lower energy), (d) band 3 (anti-symmetric band with higher energy), and (e) band 4 (symmetric band with higher energy) at $k_j = 0$. The size of yellowish glow around lattice sites qualitatively represents the intensity of the eigenstate on lattice sites, and the brightness (light yellow or dark yellow) represent the plus (light)/minus (dark) sign of the eigenstate intensity. $-2 \leq m, n \leq 3$ and $R = 2$ in (b)–(e) for illustration purposes.

We next perform numerical simulations to validate this newly observed interface mode. Eq. (29) is solved with $R = 6$, and the evolutions of the average boson number distribution along the 1D SSH chain are calculated by Eqs. (24)–(26). We first consider the case that sites $C_0$ and $D_6$ are excited by a pair of separate bosons at $\Delta \varepsilon = g$ and the lattice is in the nontrivial, nonlinear phase ($g_1 = g$, $g_2 = 3g$, $U = 2g$). In the corresponding plot of the evolution of the average boson number distribution in Fig. 7(a), one can see that the interface mode is excited and preserved throughout the simulation, as the distance between the two bosons keeps stationary. The boson scattering in Fig. 7(a) is negligible because the bands of interface modes are relatively flat, so the group velocity of the wavepacket is small. In Fig. 7(b), we take the case that $C_0$ and $D_6$ are excited but the lattice is in the trivial, nonlinear phase ($g_1 = 3g$, $g_2 = g$, $U = 2g$). There is no interface mode observed. In Fig. 7(c), the lattice is in the nontrivial, nonlinear phase, but two bosons are excited simultaneously on site $C_0$, and thus bulk modes are excited instead. Finally, in Fig. 7(d), $C_0$ and $D_6$ are excited and the lattice is in the nontrivial phase but $U = 0$, and no interface mode is found as expected. The simulations here prove that, for the Hamiltonian in Eq. (27), the nonzero onsite potential $U$, the topologically nontrivial phase ($g_2 > g_1$), and an initial excitation on the linear–nonlinear interface are all necessary conditions to successfully excite an interface mode in Fig. 5(c).

The successful demonstration of the interface modes exhibits the capability of our approach, with the two-boson synthetic lattice and the band structure analysis, to predict new physical phenomena. We also note that several previous studies used the modified Bethe ansatz and gave explicit solutions to some specific Hamiltonians with short-range interactions [30,33,35-37,40], yet these methods still face challenges when the interactions are arbitrarily nonlocal ($R \geq 2$) or the total boson number is large ($N \geq 3$). Our theoretical approach proposed here is more general and, as we show in Sections 4 and 5, is able to deal with multi-boson dynamics with long-range interactions and discover exotic states in these systems.

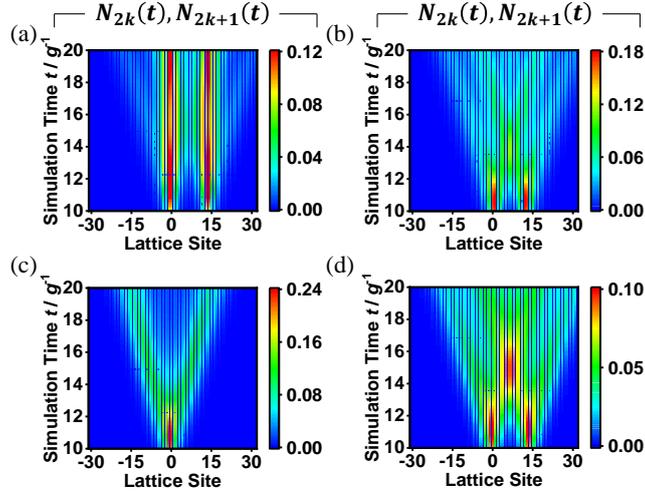

**Fig. 7. Average boson number simulation results of interface modes in the two-boson SSH synthetic lattice with interactions.** (a) Excite two bosons on sites $C_0$ and $D_6$ respectively, with $U = 2g$, $g_1 = g$, $g_2 = 3g$ and $\Delta\varepsilon = g$. (b) Excite two bosons on sites $C_0$ and $D_6$ respectively, with $U = 2g$, $g_1 = 3g$, $g_2 = g$ and $\Delta\varepsilon = g$. (c) Excite two bosons on site $C_0$ simultaneously, with $U = 2g$, $g_1 = g$, $g_2 = 3g$ and $\Delta\varepsilon = g$. (d) Excite two bosons on sites $C_0$ and $D_6$ respectively, with $U = 0$, $g_1 = g$, $g_2 = 3g$ and $\Delta\varepsilon = 0$. $R = 6$, $t_0 = 10g^{-1}$ and $\tau^2 = 10g^{-2}$ in this figure.

## 5. Three bosons on a tight-binding lattice with interactions

We have discussed the 2D synthetic SSH lattice induced by the two-boson physics in Sections 3 and 4. In this section, we demonstrate a three-dimensional (3D), three-boson synthetic lattice, which is constructed by three indistinguishable bosons on a 1D tight-binding extended Bose–Hubbard Hamiltonian, and show that it possesses exotic trimer states.

We consider a Hamiltonian which goes beyond previous studies [62-64] with nonlocal interactions:

$$H = g \sum_{k=-\infty}^{+\infty} \left( b_k^\dagger b_{k+1} + b_{k+1}^\dagger b_k \right) + \frac{U}{2} \sum_{k=-\infty}^{+\infty} \sum_{s=-R}^{R} b_{k+s}^\dagger b_k^\dagger b_k b_{k+s} , \qquad (30)$$

and the three-boson state

$$|\phi(t)\rangle = \frac{1}{\sqrt{6}} \sum_{m=-\infty}^{+\infty} \sum_{n=-\infty}^{+\infty} \sum_{p=-\infty}^{+\infty} v_{mnp}(t) b_m^\dagger b_n^\dagger b_p^\dagger |0\rangle, \quad (31)$$

By substituting Eqs. (30) and (31) into the Schrödinger equation, we get the differential equations for $v_{mnp}$'s ($R \geq 2$):

$$i \sum_{\text{permu}} \dot{v}_{mnp} = \sum_{\text{permu}} \left[ g\left(v_{m-1,n,p} + v_{m+1,n,p} + v_{m,n-1,p} + v_{m,n+1,p} + v_{m,n,p-1} + v_{m,n,p+1}\right) + \left(\delta_{|m-n|\leq R} + \delta_{|n-p|\leq R} + \delta_{|p-m|\leq R}\right) U v_{mnp} \right]. \quad (32)$$

For this three-boson case, the wavefunction exchange symmetry is defined as $v_{mnp} = v_{mpn} = v_{nmp} = v_{npm} = v_{pmn} = v_{pnm}$, and then the permutation sum in Eq. (32) can be removed:

$$i\dot{v}_{mnp} = g\left(v_{m-1,n,p} + v_{m+1,n,p} + v_{m,n-1,p} + v_{m,n+1,p} + v_{m,n,p-1} + v_{m,n,p+1}\right) + \left(\delta_{|m-n|\leq R} + \delta_{|n-p|\leq R} + \delta_{|p-m|\leq R}\right) U v_{mnp}. \quad (33)$$

The first six terms in Eq. (33) correspond to a 3D tight-binding lattice, which spans the full 3D $(m,n,p)$ space with equal coupling strengths $g$, as is shown by the cubic structure in Fig. 8(a). (If we do not assume the exchange symmetry, and instead restrict that $v_{mnp} = 0$ except in the region $m \leq n \leq p$, Eq. (32) would be translated to the tetrahedron structure in Fig. S1(c) [49], which does not support the following band structure analysis.) After ignoring the nonlinearity for the moment (i.e. $U=0$), the band structure of the cubic three-boson synthetic lattice is

$$\varepsilon_{\text{tight-binding}}^{3D}(k_m, k_n, k_p) = 2g\left(\cos k_m + \cos k_n + \cos k_p\right), \quad (34)$$

which can be shown by the energy isosurfaces in Fig. 8(b). At the center point of the cubic reciprocal lattice ($\Gamma$ point), $\varepsilon$ takes its maximum value of $6g$, while at corners (R point), $\varepsilon$ takes the minimum value of $-6g$. Now, we consider the added nonzero onsite potential $U$ on lattice sites around the line $m = n = p$ (the $j$ axis, $\hat{\mathbf{j}} = (\hat{\mathbf{m}} + \hat{\mathbf{n}} + \hat{\mathbf{p}})/\sqrt{3}$), as indicated by the green spots in Fig. 8(a). One can perform Fourier transform along the $j$ axis and obtain the projected band structure $\varepsilon(k_j)$. Here $k_j = k_m + k_n + k_p$. For the linear case with $U=0$, $\varepsilon(k_j)$ can be understood from Eq. (34) by a projection onto the $\Gamma$–R direction and then band folding into the first Brillouin zone (see Figs. S4 (e) and (f) in the Supplemental Material [49]).

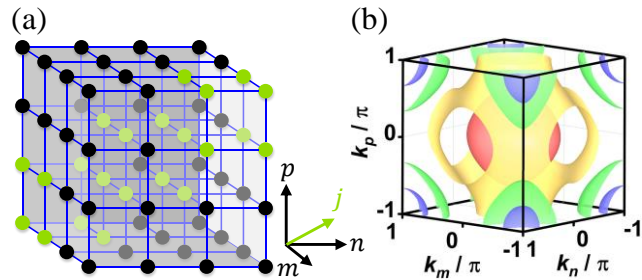

**Fig. 8. The three-boson tight-binding synthetic lattice with interactions.** (a) The three-boson synthetic lattice with the exchange symmetry restriction on the wavefunction $v_{mnp}(t)$, where green spots indicate nonzero onsite potentials. (b) Energy isosurface of the 3D band structure of the lattice in (a) in the first Brillouin zone when $U=0$. Red: $\varepsilon = 3g$; yellow: $\varepsilon = 0$; green: $\varepsilon = -3g$; blue: $\varepsilon = -5g$.

We then consider the strong-interaction case with $U = 12g$ and plot the projected band structure $\varepsilon(k_j)$ in Fig. 9(a). One can see that there are four separate quasi-continuum bands. We find that these bands correspond to different multi-boson states when we plot the eigenstates of these four quasi-continuum bands in the $l_1 - l_2$ plane, as shown by the diagram in Fig. 9(b). Here $\hat{\mathbf{l}}_1 = (\hat{\mathbf{n}} - \hat{\mathbf{m}})/\sqrt{2}$ and $\hat{\mathbf{l}}_2 = (\hat{\mathbf{p}} - \hat{\mathbf{m}})/\sqrt{2}$ are two linearly independent directions orthogonal to the $j$ axis in the 3D space, and $|l_1|$, $|l_2|$, $|l_1 - l_2|$ are actually the relative distances between each two of these three bosons. The $l_1 - l_2$ plane is divided into 19 regions by 6 lines $|l_1| = \pm R$, $|l_2| = \pm R$, $|l_1 - l_2| = \pm R$, classified into 4 categories, which are dominated by eigenstates from different bands. In the blue regions as shown in Fig. 9(b), we find that they correspond to the band with lowest energy in Fig. 9(a), as confirmed by the eigenstate distribution at $(k_j, \varepsilon) = (0, 5.62g)$ shown in Fig. 9(c). The areas of the blue regions are infinitely large if $-\infty < l_1, l_2 < +\infty$, and none of the three $\delta$-functions in Eq. (33) equals to one in these regions. It should also be noted that, although in Fig. 9(c), only the left-top and right-bottom corners of the $l_1 - l_2$ plane are occupied, the eigenstate distributions at other eigen-energies in the lowest-energy band fill up the rest of the blue regions in Fig. 9(b). These states are three-boson scattering states where bosons are not bound to each other and do not interact with each other via the nonlinearity. In the infinite cyan regions, one of the three $\delta$-functions in Eq. (33) equals to one, and these states consist of a doublon and a free particle (see the eigenstate distribution at $(k_j, \varepsilon) = (0, 17.84g)$ shown in Fig. 9(d)), residing in the second-lowest energy band in Fig. 9(a). Moreover, the red region denotes the tightly-bound triplon states (see the eigenstate distribution at $(k_j, \varepsilon) = (0, 41.79g)$ shown in Fig. 9(f)) with the strongest interaction (consequently the highest energy) in Fig. 9(a). Nevertheless, the most intriguing result discovered here is the hexagram-like, *weakly-bound triplon state* (see the eigenstate distribution at $(k_j, \varepsilon) = (0, 28.85g)$ shown in Fig. 9(e)) in the orange regions. Such an exotic state in Fig. 9(e) has not been thoroughly understood previously in the model without the nonlocal interaction [62]. In this state, compared to the tightly-bound trimer in the red region, the relative distances between each two of these three particles cannot be all smaller than $R$, while compared to the dimer–monomer states in the cyan regions, none of the relative distances between each two of these three particles is allowed to be larger than $2R$, albeit there is no artificial discontinuity at $2R$ in the Hamiltonian in Eq. (30). This 'virtue potential wall' at $2R$ is entirely attributed to the boson–boson blockade effect. Therefore, in this fascinating state in Fig. 9(e), three bosons are loosely localized together, neither too close to nor too far from one another.

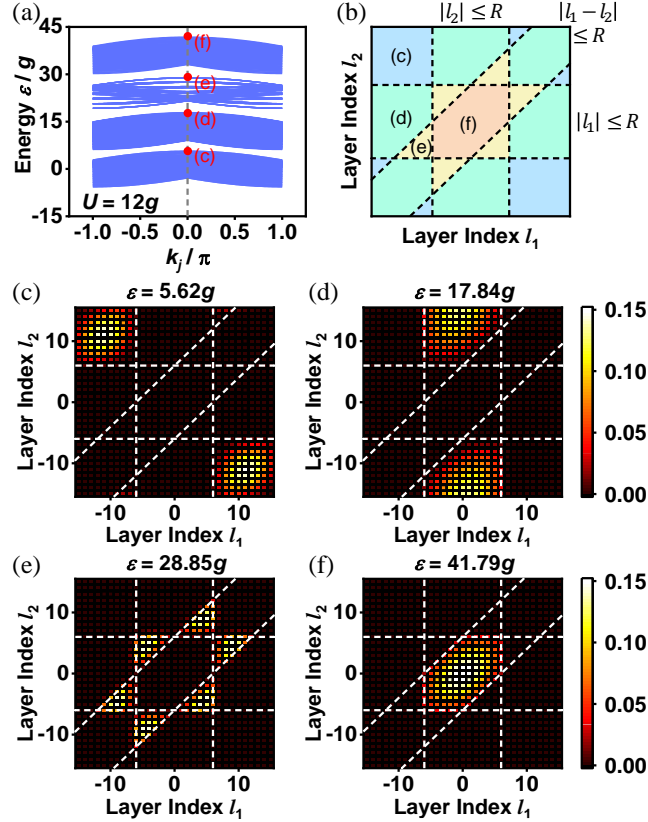

**Fig. 9. Projected band structure and eigenstate distributions of the three-boson tight-binding synthetic lattice with interactions.** (a) Projected band structure of the three-boson synthetic lattice with $U = 12g$. (b) Regional diagram of eigenstate distributions. Blue area: bands of lowest energy; cyan area: bands of second-lowest energy; orange area: bands of second-highest energy; red area: bands of highest-energy. (c)–(f) Representative eigenstate distributions corresponding to the band structure at $k_j = 0$ (grey dash in (a)) and at (c) $\varepsilon = 5.62g$, (d) $\varepsilon = 17.84g$, (e) $\varepsilon = 28.85g$, (f) $\varepsilon = 41.79g$. $R = 6$ and $-15 \leq l_1, l_2 \leq 15$ in this figure.

## 6. Discussions and conclusions

In this paper, we propose a theoretical framework to treat the multi-boson dynamics in a one-dimensional lattice as an *N*-boson synthetic lattice by applying exchange symmetry restrictions to the wavefunction. In such an artificially constructed *N*-boson synthetic lattice, one can mathematically perform band structure analysis and the resulting *N*-dimensional band structure provides a novel perspective to analyze the multi-boson dynamics on 1D lattices. We show that, for complicated Hamiltonians with boson–boson interactions, projected band structures could be understood from the full *N*-dimensional band structure, and nontrivial multi-boson states (such as interface modes in Section 4 and weakly-bound triplon states in Section 5) can be successfully predicted. In addition, the 1D lattice required in this paper to hold these multiple bosons, can be in either real or synthetic dimensions *per se*. If this 1D lattice itself is in synthetic dimensions (e.g. formed by exploiting boson frequencies, orbital angular momenta, etc. [13]), our theory suggests that this multi-boson approach might enable the construction of arbitrarily multi-dimensional systems on an otherwise zero-dimensional platform, together with conventional strategies of synthetic dimensions. Mathematically, if *N* bosons are excited on a platform with *D* real dimensions and *M* synthetic dimensions, the system can be eventually expanded to $(D+M) \times N$ dimensions (with exchange symmetry restrictions).

We also briefly suggest several possible experimental platforms for exploring *N*-boson dynamics in the 1D lattices where our proposed approach can be conducted. For lattices in the real space, coupled photonic waveguides and cavities [42,43,65-71], cold bosonic atoms in optical lattices [72-75], and superconducting circuits [76-79] are state-of-the-art technologies for implementations. For example, the coupled waveguide array fabricated to study the two-photon quantum walks in a 1D SSH lattice [71] gives an experimental demonstration of our discussions in Section 3. As for the synthetic space, 1D tight-binding model and SSH Hamiltonians without interactions are readily accessible via synthetic frequency dimensions in few ring resonators with electro-optic modulation [15,57,80-82]. Artificial boundaries of the lattices along frequency dimensions could be achieved by incorporating group velocity dispersions in ring resonators or introducing lossy absorbers at particular frequencies [70,83], and interactive terms in Hamiltonians can be constructed from four-wave-mixing processes with the nonlinear effect [84]. Additionally, arbitrarily long-range coupling in synthetic frequency dimensions is also possible by special designs of the electro-optic modulation profile [85].

In conclusion, we systematically develop a theoretical framework on the creation of the *N*-boson synthetic lattice that enlightens insights into the studies of both synthetic dimensions and the physics of multiple interactive indistinguishable bosons. The connection between multiple bosons and multiple dimensions is highlighted. Through band structure analysis, novel dynamics of multi-boson states are unveiled, and can be confirmed by numerical simulations. Our study hence points out a new way towards the studies of boson–boson interactions and multi-boson dynamics on lattices, and also holds potentials for exploring important multi-boson manipulations together with nonlinearity, nontrivial topology, and/or boson entanglements with possible applications in fields of quantum computations, quantum simulations, and quantum information processing.

**Acknowledgements**

The research is supported by National Natural Science Foundation of China (11974245), National Key R&D Program of China (2017YFA0303701), Shanghai Municipal Science and Technology Major Project (2019SHZDZX01), and Natural Science Foundation of Shanghai (19ZR1475700). L. Y. acknowledges support from the Program for Professor of Special Appointment (Eastern Scholar) at Shanghai Institutions of Higher Learning. X. C. also acknowledges the support from Shandong Quancheng Scholarship (00242019024).

**References**

[1] A. P. Schnyder, S. Ryu, A. Furusaki, and A. W. W. Ludwig, *Classification of Topological Insulators and Superconductors in Three Spatial Dimensions*, Phys. Rev. B **78**, 195125 (2008).

[2] A. Kitaev, *Periodic Table for Topological Insulators and Superconductors*, AIP Conf. Proc. **1134**, 22 (2009).

[3] S. Ryu, A. P. Schnyder, A. Furusaki, and A. W. W. Ludwig, *Topological Insulators and Superconductors: Tenfold Way and Dimensional Hierarchy*, New J. Phys. **12**, 065010 (2010).

[4] M. Z. Hasan and C. L. Kane, *Colloquium: Topological Insulators*, Rev. Mod. Phys. **82**, 3045 (2010).

[5] S. T. Ruggiero, T. W. Barbee, Jr., and M. R. Beasley, *Superconductivity in Quasi-Two-Dimensional Layered Composites*, Phys. Rev. Lett. **45**, 1299 (1980).

[6] Y. Guo, Y.-F. Zhang, X.-Y. Bao, T.-Z. Han, Z. Tang, L.-X. Zhang, W.-G. Zhu, E. G. Wang, Q. Niu, Z. Q. Qiu *et al.*, *Superconductivity Modulated by Quantum Size Effects*, Science **306**, 1915 (2004).

[7] K. Y. Arutyunov, D. S. Golubev, and A. D. Zaikin, *Superconductivity in One Dimension*, Phys. Rep. **464**, 1 (2008).

[8] S. Qin, J. Kim, Q. Niu, and C.-K. Shih, *Superconductivity at the Two-Dimensional Limit*, Science **324**, 1314 (2009).

[9] J. Guo, Y. Zhou, C. Huang, S. Cai, Y. Sheng, G. Gu, C. Yang, G. Lin, K. Yang, A. Li *et al.*, *Crossover from Two-Dimensional to Three-Dimensional Superconducting States in Bismuth-Based Cuprate Superconductor*, Nat. Phys. **16**, 295 (2020).

[10] A. Celi, P. Massignan, J. Ruseckas, N. Goldman, I. B. Spielman, G. Juzeliūnas, and M. Lewenstein, *Synthetic Gauge Fields in Synthetic Dimensions*, Phys. Rev. Lett. **112**, 043001 (2014).

[11] M. Mancini, G. Pagano, G. Cappellini, L. Livi, M. Rider, J. Catani, C. Sias, P. Zoller, M. Inguscio, M. Dalmonte *et al.*, *Observation of Chiral Edge States with Neutral Fermions in Synthetic Hall Ribbons*, Science **349**, 1510 (2015).

[12] T. Ozawa, H. M. Price, N. Goldman, O. Zilberberg, and I. Carusotto, *Synthetic Dimensions in Integrated Photonics: From Optical Isolation to Four-Dimensional Quantum Hall Physics*, Phys. Rev. A **93**, 043827 (2016).

[13] L. Yuan, Q. Lin, M. Xiao, and S. Fan, *Synthetic Dimension in Photonics*, Optica **5**, 1396 (2018).

[14] E. Lustig, S. Weimann, Y. Plotnik, Y. Lumer, M. A. Bandres, A. Szameit, and M. Segev, *Photonic Topological Insulator in Synthetic Dimensions*, Nature (London) **567**, 356 (2019).

[15] A. Dutt, Q. Lin, L. Yuan, M. Minkov, M. Xiao, and S. Fan, *A Single Photonic Cavity with Two Independent Physical Synthetic Dimensions*, Science **367**, 59 (2020).

[16] Z. Liu, Q. Zhang, F. Qin, D. Zhang, X. Liu, and J. J. Xiao, *Surface States Ensured by a Synthetic Weyl Point in One-Dimensional Plasmonic Dielectric Crystals with Broken Inversion Symmetry*, Phys. Rev. B **99**, 085441 (2019).

[17]     Z. Liu, Q. Zhang, F. Qin, D. Zhang, X. Liu, and J. J. Xiao, *Type-II Dirac Point and Extreme Dispersion in One-Dimensional Plasmonic-Dielectric Crystals with Off-Axis Propagation*, Phys. Rev. A **99**, 043828 (2019).

[18]     T.-S. Zeng, C. Wang, and H. Zhai, *Charge Pumping of Interacting Fermion Atoms in the Synthetic Dimension*, Phys. Rev. Lett. **115**, 095302 (2015).

[19]     L. F. Livi, G. Cappellini, M. Diem, L. Franchi, C. Clivati, M. Frittelli, F. Levi, D. Calonico, J. Catani, M. Inguscio *et al.*, *Synthetic Dimensions and Spin-Orbit Coupling with an Optical Clock Transition*, Phys. Rev. Lett. **117**, 220401 (2016).

[20]     H. M. Price, T. Ozawa, and H. Schomerus, *Synthetic Dimensions and Topological Chiral Currents in Mesoscopic Rings*, Phys. Rev. Res. **2**, 032017(R) (2020).

[21]     J. Wang, J.-Y. Yang, I. M. Fazal, N. Ahmed, Y. Yan, H. Huang, Y. Ren, Y. Yue, S. Dolinar, M. Tur *et al.*, *Terabit Free-Space Data Transmission Employing Orbital Angular Momentum Multiplexing*, Nat. Photonics **6**, 488 (2012).

[22]     M. Mirhosseini, O. S. Magaña-Loaiza, M. N. O'Sullivan, B. Rodenburg, M. Malik, M. P. J. Lavery, M. J. Padgett, D. J. Gauthier, and R. W. Boyd, *High-Dimensional Quantum Cryptography with Twisted Light*, New J. Phys. **17**, 033033 (2015).

[23]     H.-H. Lu, J. M. Lukens, N. A. Peters, O. D. Odele, D. E. Leaird, A. M. Weiner, and P. Lougovski, *Electro-Optic Frequency Beam Splitters and Tritters for High-Fidelity Photonic Quantum Information Processing*, Phys. Rev. Lett. **120**, 030502 (2018).

[24]     C. Joshi, A. Farsi, S. Clemmen, S. Ramelow, and A. L. Gaeta, *Frequency Multiplexing for Quasi-Deterministic Heralded Single-Photon Sources*, Nat. Commun. **9**, 847 (2018).

[25]     S. Buddhiraju, A. Dutt, M. Minkov, I. A. D. Williamson, and S. Fan, *Arbitrary Linear Transformations for Photons in the Frequency Synthetic Dimension*, arXiv:2009.02008.

[26]     A. Aspuru-Guzik and P. Walther, *Photonic Quantum Simulators*, Nat. Phys. **8**, 285 (2012).

[27]     A. Schreiber, A. Gábris, P. P. Rohde, K. Laiho, M. Štefaňák, V. Potoček, C. Hamilton, I. Jex, and C. Silberhorn, *A 2D Quantum Walk Simulation of Two-Particle Dynamics*, Science **336**, 55 (2012).

[28]     T. Sturges, T. McDermott, A. Buraczewski, W. R. Clements, J. J. Renema, S. W. Nam, T. Gerrits, A. Lita, W. S. Kolthammer, A. Eckstein *et al.*, *Quantum Simulations with Multiphoton Fock States*, arXiv:1906.00678.

[29]     K. Winkler, G. Thalhammer, F. Lang, R. Grimm, J. Hecker Denschlag, A. J. Daley, A. Kantian, H. P. Büchler, and P. Zoller, *Repulsively Bound Atom Pairs in an Optical Lattice*, Nature (London) **441**, 853 (2006).

[30]     M. Valiente and D. Petrosyan, *Two-Particle States in the Hubbard Model*, J. Phys. B: At. Mol. Opt. Phys. **41**, 161002 (2008).

[31] M. Valiente and D. Petrosyan, *Quantum Dynamics of One and Two Bosonic Atoms in a Combined Tight-Binding Periodic and Weak Parabolic Potential*, EPL **83**, 30007 (2008).

[32] S. Longhi and G. Della Valle, *Tamm–Hubbard Surface States in the Continuum*, J. Phys.: Condens. Matter **25**, 235601 (2013).

[33] M. Valiente and D. Petrosyan, *Scattering Resonances and Two-Particle Bound States of the Extended Hubbard Model*, J. Phys. B: At. Mol. Opt. Phys. **42**, 121001 (2009).

[34] M. Valiente, *Lattice Two-Body Problem with Arbitrary Finite-Range Interactions*, Phys. Rev. A **81**, 042102 (2010).

[35] M. A. Gorlach and A. N. Poddubny, *Interaction-Induced Two-Photon Edge States in an Extended Hubbard Model Realized in a Cavity Array*, Phys. Rev. A **95**, 033831 (2017).

[36] M. Di Liberto, A. Recati, I. Carusotto, and C. Menotti, *Two-Body Physics in the Su-Schrieffer-Heeger Model*, Phys. Rev. A **94**, 062704 (2016).

[37] M. A. Gorlach and A. N. Poddubny, *Topological Edge States of Bound Photon Pairs*, Phys. Rev. A **95**, 053866 (2017).

[38] M. Di Liberto, A. Recati, I. Carusotto, and C. Menotti, *Two-Body Bound and Edge States in the Extended SSH Bose-Hubbard Model*, Eur. Phys. J. Spec. Top. **226**, 2751 (2017).

[39] A. M. Marques and R. G. Dias, *Topological Bound States in Interacting Su–Schrieffer–Heeger Rings*, J. Phys.: Condens. Matter **30**, 305601 (2018).

[40] A. A. Stepanenko and M. A. Gorlach, *Interaction-Induced Topological States of Photon Pairs*, Phys. Rev. A **102**, 013510 (2020).

[41] R. Khomeriki, D. O. Krimer, M. Haque, and S. Flach, *Interaction-Induced Fractional Bloch and Tunneling Oscillations*, Phys. Rev. A **81**, 065601 (2010).

[42] S. Longhi, *Photonic Bloch Oscillations of Correlated Particles*, Opt. Lett. **36**, 3248 (2011).

[43] G. Corrielli, A. Crespi, G. Della Valle, S. Longhi, and R. Osellame, *Fractional Bloch Oscillations in Photonic Lattices*, Nat. Commun. **4**, 1555 (2013).

[44] S.-y. Chu and D.-c. Su, *Operation of a Two-Mode Laser in a Three-Level Atomic System with a Common Upper Level*, Phys. Rev. A **25**, 3169 (1982).

[45] D.-W. Wang, H. Cai, R.-B. Liu, and M. O. Scully, *Mesoscopic Superposition States Generated by Synthetic Spin-Orbit Interaction in Fock-State Lattices*, Phys. Rev. Lett. **116**, 220502 (2016).

[46] K. Tschernig, R. de J. León-Montiel, A. Pérez-Leija, and K. Busch, *Multiphoton Synthetic Lattices in Multiport Waveguide Arrays: Synthetic Atoms and Fock Graphs*, Photonics Res. **8**, 1161 (2020).

[47] H. Cai and D.-W. Wang, *Topological Phases of Quantized Light*, Natl. Sci. Rev. **8**, nwaa196 (2021).

[48]     A. Yariv, *Coupled-Mode Theory for Guided-Wave Optics*, IEEE J. Quantum Electron. **9**, 919 (1973).

[49]     Supplemental Material.

[50]     A. P. Sutton, M. W. Finnis, D. G. Pettifor, and Y. Ohta, *The Tight-Binding Bond Model*, J. Phys. C: Solid State Phys. **21**, 35 (1988).

[51]     S. Fan, Ş. E. Kocabaş, and J.-T. Shen, *Input-Output Formalism for Few-Photon Transport in One-Dimensional Nanophotonic Waveguides Coupled to a Qubit*, Phys. Rev. A **82**, 063821 (2010).

[52]     W. P. Su, J. R. Schrieffer, and A. J. Heeger, *Solitons in Polyacetylene*, Phys. Rev. Lett. **42**, 1698 (1979).

[53]     E. J. Meier, F. A. An, and B. Gadway, *Observation of the Topological Soliton State in the Su–Schrieffer–Heeger Model*, Nat. Commun. **7**, 13986 (2016).

[54]     S. Liu, W. Gao, Q. Zhang, S. Ma, L. Zhang, C. Liu, Y. J. Xiang, T. J. Cui, and S. Zhang, *Topologically Protected Edge State in Two-Dimensional Su–Schrieffer–Heeger Circuit*, Research **2019**, 8609875 (2019).

[55]     L.-Y. Zheng, V. Achilleos, O. Richoux, G. Theocharis, and V. Pagneux, *Observation of Edge Waves in a Two-Dimensional Su-Schrieffer-Heeger Acoustic Network*, Phys. Rev. Appl. **12**, 034014 (2019).

[56]     W. Zhang and X. Zhang, *Photonic Quadrupole Topological Phases in Zero-Dimensional Cavity with Synthetic Dimensions*, arXiv:1906.02967.

[57]     A. Dutt, M. Minkov, I. A. D. Williamson, and S. Fan, *Higher-Order Topological Insulators in Synthetic Dimensions*, Light Sci. Appl. **9**, 131 (2020).

[58]     M. Kim and J. Rho, *Topological Edge and Corner States in a Two-Dimensional Photonic Su-Schrieffer-Heeger Lattice*, Nanophotonics **9**, 3227 (2020).

[59]     F. Liu and K. Wakabayashi, *Novel Topological Phase with a Zero Berry Curvature*, Phys. Rev. Lett. **118**, 076803 (2017).

[60]     D. Obana, F. Liu, and K. Wakabayashi, *Topological Edge States in the Su-Schrieffer-Heeger Model*, Phys. Rev. B **100**, 075437 (2019).

[61]     A. Coutant, V. Achilleos, O. Richoux, G. Theocharis, and V. Pagneux, *Robustness of Topological Corner Modes against Disorder with Application to Acoustic Networks*, Phys. Rev. B **102**, 214204 (2020).

[62]     M. Valiente, D. Petrosyan, and A. Saenz, *Three-Body Bound States in a Lattice*, Phys. Rev. A **81**, 011601(R) (2010).

[63]     A. Miranowicz, M. Paprzycka, Y.-x. Liu, J. Bajer, and F. Nori, *Two-Photon and Three-Photon Blockades in Driven Nonlinear Systems*, Phys. Rev. A **87**, 023809 (2013).

[64] Q.-Y. Liang, A. V. Venkatramani, S. H. Cantu, T. L. Nicholson, M. J. Gullans, A. V. Gorshkov, J. D. Thompson, C. Chin, M. D. Lukin, and V. Vuletić, *Observation of Three-Photon Bound States in a Quantum Nonlinear Medium*, Science **359**, 783 (2018).

[65] A. Yariv, Y. Xu, R. K. Lee, and A. Scherer, *Coupled-Resonator Optical Waveguide: A Proposal and Analysis*, Opt. Lett. **24**, 711 (1999).

[66] M. J. Hartmann, F. G. S. L. Brandão, and M. B. Plenio, *Strongly Interacting Polaritons in Coupled Arrays of Cavities*, Nat. Phys. **2**, 849 (2006).

[67] M. J. Hartmann, F. G. S. L. Brandão, and M. B. Plenio, *Quantum Many-Body Phenomena in Coupled Cavity Arrays*, Laser Photonics Rev. **2**, 527 (2008).

[68] R. O. Umucalılar and I. Carusotto, *Artificial Gauge Field for Photons in Coupled Cavity Arrays*, Phys. Rev. A **84**, 043804 (2011).

[69] D. O. Krimer and R. Khomeriki, *Realization of Discrete Quantum Billiards in a Two-Dimensional Optical Lattice*, Phys. Rev. A **84**, 041807(R) (2011).

[70] L. Yuan, Y. Shi, and S. Fan, *Photonic Gauge Potential in a System with a Synthetic Frequency Dimension*, Opt. Lett. **41**, 741 (2016).

[71] F. Klauck, M. Heinrich, and A. Szameit, *Photonic Two-Particle Quantum Walks in Su–Schrieffer–Heeger Lattices*, Photonics Res. **9**, A1 (2021).

[72] D. Jaksch, C. Bruder, J. I. Cirac, C. W. Gardiner, and P. Zoller, *Cold Bosonic Atoms in Optical Lattices*, Phys. Rev. Lett. **81**, 3108 (1998).

[73] R. Grimm, M. Weidemüller, and Y. B. Ovchinnikov, *Optical Dipole Traps for Neutral Atoms*, Adv. At. Mol. Opt. Phys. **42**, 95 (2000).

[74] I. Bloch, J. Dalibard, and W. Zwerger, *Many-Body Physics with Ultracold Gases*, Rev. Mod. Phys. **80**, 885 (2008).

[75] C. Chin, R. Grimm, P. Julienne, and E. Tiesinga, *Feshbach Resonances in Ultracold Gases*, Rev. Mod. Phys. **82**, 1225 (2010).

[76] M. H. Devoret and R. J. Schoelkopf, *Superconducting Circuits for Quantum Information: An Outlook*, Science **339**, 1169 (2013).

[77] X. Gu, S. Chen, and Y.-x. Liu, *Topological Edge States and Pumping in a Chain of Coupled Superconducting Qubits*, arXiv:1711.06829.

[78] Z. Yan, Y.-R. Zhang, M. Gong, Y. Wu, Y. Zheng, S. Li, C. Wang, F. Liang, J. Lin, Y. Xu *et al.*, *Strongly Correlated Quantum Walks with a 12-Qubit Superconducting Processor*, Science **364**, 753 (2019).

[79] Y. Yanay, J. Braumüller, S. Gustavsson, W. D. Oliver, and C. Tahan, *Two-Dimensional Hard-Core Bose–Hubbard Model with Superconducting Qubits*, npj Quantum Inf. **6**, 58 (2020).

[80] L. Yuan and S. Fan, *Bloch Oscillation and Unidirectional Translation of Frequency in a Dynamically Modulated Ring Resonator*, Optica **3**, 1014 (2016).

[81] A. Dutt, M. Minkov, Q. Lin, L. Yuan, D. A. B. Miller, and S. Fan, *Experimental Band Structure Spectroscopy along a Synthetic Dimension*, Nat. Commun. **10**, 3122 (2019).

[82] G. Li, Y. Zheng, A. Dutt, D. Yu, Q. Shan, S. Liu, L. Yuan, S. Fan, and X. Chen, *Dynamic Band Structure Measurement in the Synthetic Space*, Sci. Adv. **7**, eabe4335 (2021).

[83] L. Yuan, Q. Lin, A. Zhang, M. Xiao, X. Chen, and S. Fan, *Photonic Gauge Potential in One Cavity with Synthetic Frequency and Orbital Angular Momentum Dimensions*, Phys. Rev. Lett. **122**, 083903 (2019).

[84] L. Yuan, A. Dutt, M. Qin, S. Fan, and X. Chen, *Creating Locally Interacting Hamiltonians in the Synthetic Frequency Dimension for Photons*, Photonics Res. **8**, B8 (2020).

[85] L. Yuan, M. Xiao, Q. Lin, and S. Fan, *Synthetic Space with Arbitrary Dimensions in a Few Rings Undergoing Dynamic Modulation*, Phys. Rev. B **97**, 104105 (2018).